\documentclass[aps,twocolumn]{revtex4-2}

\usepackage[T1]{fontenc}
\usepackage{charter}

\usepackage{amsmath,amssymb,amsfonts,color,graphicx,tabularx}

\usepackage[unicode=true,colorlinks=true]{hyperref}

\hypersetup{linkcolor=blue,citecolor=blue,urlcolor=blue}

\begin{document}

\title{Non-Abelian fractional quantum Hall states at filling factor $3/4$}

\author{Kai-Wen Huang}
\affiliation{School of Physics and Wuhan National High Magnetic Field Center, Huazhong University of Science and Technology, Wuhan 430074, China}

\author{Ying-Hai Wu}
\email{yinghaiwu88@hust.edu.cn}
\affiliation{School of Physics and Wuhan National High Magnetic Field Center, Huazhong University of Science and Technology, Wuhan 430074, China}

\begin{abstract}
Fractional quantum Hall states have been observed at filling factor $\nu=3/4$ in GaAs hole system and bilayer graphene. In theoretical bootstrap analysis, it was revealed that non-Abelian topological orders with Ising anyons can be realized at $\nu=3/4$, which exhibit $12$ fold ground state degeneracy on the torus. The properties of $\nu=3/4$ states can be analyzed using two complementary approaches. In the first one, they are treated as particle-hole conjugate of $\nu=1/4$ Moore-Read types states. In the second one, they are mapped to composite fermions with reverse flux attachment at effective filling factor $3/2$, whose integral part realizes an integer quantum Hall state and the fractional part realizes $\nu=1/2$ Moore-Read type states. For bilayer graphene with appropriate Landau level mixing, numerical calculations found $12$ quasi-degenerate ground states on the torus at $\nu=3/4$. Chiral graviton spectral functions of these states have one low energy peak with negative chirality and one high energy peak with positive chirality. This points to a specific member of the Moore-Read type states and agrees with the deduction based on daughter states.
\end{abstract}

\maketitle

\section{Introduction} 

Fractional quantum Hall states constitute an prominent example of strongly correlated topological phases~\cite{Halperin-Book}. In the most common setting, electrons are confined to two dimensional space and exposed to a perpendicular magnetic field such that dispersionless Landau levels (LLs) arise. An intricate interplay between the single-particle states and electron correlations stabilize many fascinating states. While experimental systems have underlying lattice structures, it is usually sufficient to employ the continuum description without worrying about lattice effects because Landau orbitals are much larger than crystalline unit cells. Thouless et al. uncovered the topological underpinning of quantum Hall physics as they linked the Hall conductance of non-interacting electrons to the Chern number of occupied energy bands~\cite{Thouless1982}. This description is independent from LLs and can be extended in various ways~\cite{Haldane1988b}. In principle, FQH states without magnetic field can be realized in energy bands that satisfy certain conditions~\cite{LiuZ2024,Jackson2015,Simon2020,WangJ2021,Ledwith2023,Andrews2024}. A few ground breaking experiments on twisted MoTe$_{2}$ and rhombohedral multilayer graphene have indeed achieved this goal~\cite{CaiJQ2023,ZengYH2023,ParkHJ2023,XuF2023,LuZG2024,XieJ2025}.

With or without magnetic fields, FQH states exhibit many remarkable universal properties that are described in the framework of topological order (TO)~\cite{WenXG1989}. It was predicted shortly after the original observation of FQH states that they posses anyonic excitations~\cite{Arovas1984} and compelling evidence has been reported in recent years~\cite{Bartolomei2020,Nakamura2020,Nakamura2023,Werkmeister2025,Samuelson2024}. The existence of non-Abelian anyons was also proposed in the context of FQH states~\cite{Moore1991,WenXG1991a} and has drawn considerable attention for their intriguing nature and potential utitlity in topological quantum computation~\cite{Kitaev2003,Nayak2008}. The Moore-Read Pfaffian state and its variants~\cite{Levin2007,LeeSS2007,SonDT2015,Zucker2016} are an important class of non-Abelian states that may be realized in multiple platforms~\cite{MaKW2024}. A dramatic manifestation of these states is that they occur at filling factors $\nu=p/q$ with denominator $q=2$ whereas most other FQH states are found at odd $q$~\cite{Willett1987}. In two experiments about GaAs hole system~\cite{WangCY2022} and bilayer graphene (BLG)~\cite{Kumar2024}, $\nu=3/4$ FQH states have been reported. Another work found resistance dips at $\nu=1/4$ in GaAs hole system, but there is a highly insulating background. This may indicate the competition of FQH state with Wigner crystals~\cite{WangCY2023a}. There was also one report about $1/2$ and $1/4$ states in monolayer graphene~\cite{Zibrov2018}. It was proposed that they may be multi-component states with unpolarized valley degrees of freedom. This scenario was examined in detail for the $1/2$ states~\cite{WuYH2022b} but no conclusive results were found at $1/4$. 

The $3/4$ states are very intriguing and call for detailed investigations. We assume that the internal degrees of freedom are polarized so multi-component states are not relevant. It has already been analyzed in the experimental papers, but we shall provide additional results about it. Based on a bootstrap analysis, strong constraints can be placed on the possible TOs at $\nu=3/4$~\cite{ChengM2025}. It was proposed that, if a system defined on the torus has $12$ degenerate ground states at $\nu=3/4$, one of four TOs with Ising anyons is realized. However, this approach does not directly yield microscopic wave functions associated with these possibilities. It is quite plausible that the two experiments have the same fundamental mechanism that can be understood in two complementary ways~\cite{WangCY2022,Kumar2024}. In the first approach, one construct $\nu=1/4$ FQH states from the usual Moore-Read type states and then perform particle-hole conjugation to obtain $3/4$ states. In the second approach, electrons are turned into composite fermions (CFs) with effective filling factor $3/2$ such that the $1/2$ part form Moore-Read type states~\cite{Jain1989a}. Numerical calculations are performed in the context of BLG to search for signatures of these states. Using a microscopic model that takes into account both the orbital characters of LLs and strong mixing between them, 12 quasi-degenerate ground states are found on the torus in a certain parameter regime. An inspection of the chiral graviton spectral functions~\cite{YangK2016,LiouSF2019,NguyenDX2021a,Haldane2021,NguyenDX2021b,NguyenDX2022,Balram2022a,WangYZ2023,LiangJH2024} indicates that this system realizes the anti-Pfaffian type TO.

\section{Non-Abelian Candidates} 

\begin{figure}[ht]
\includegraphics[width=0.48\textwidth]{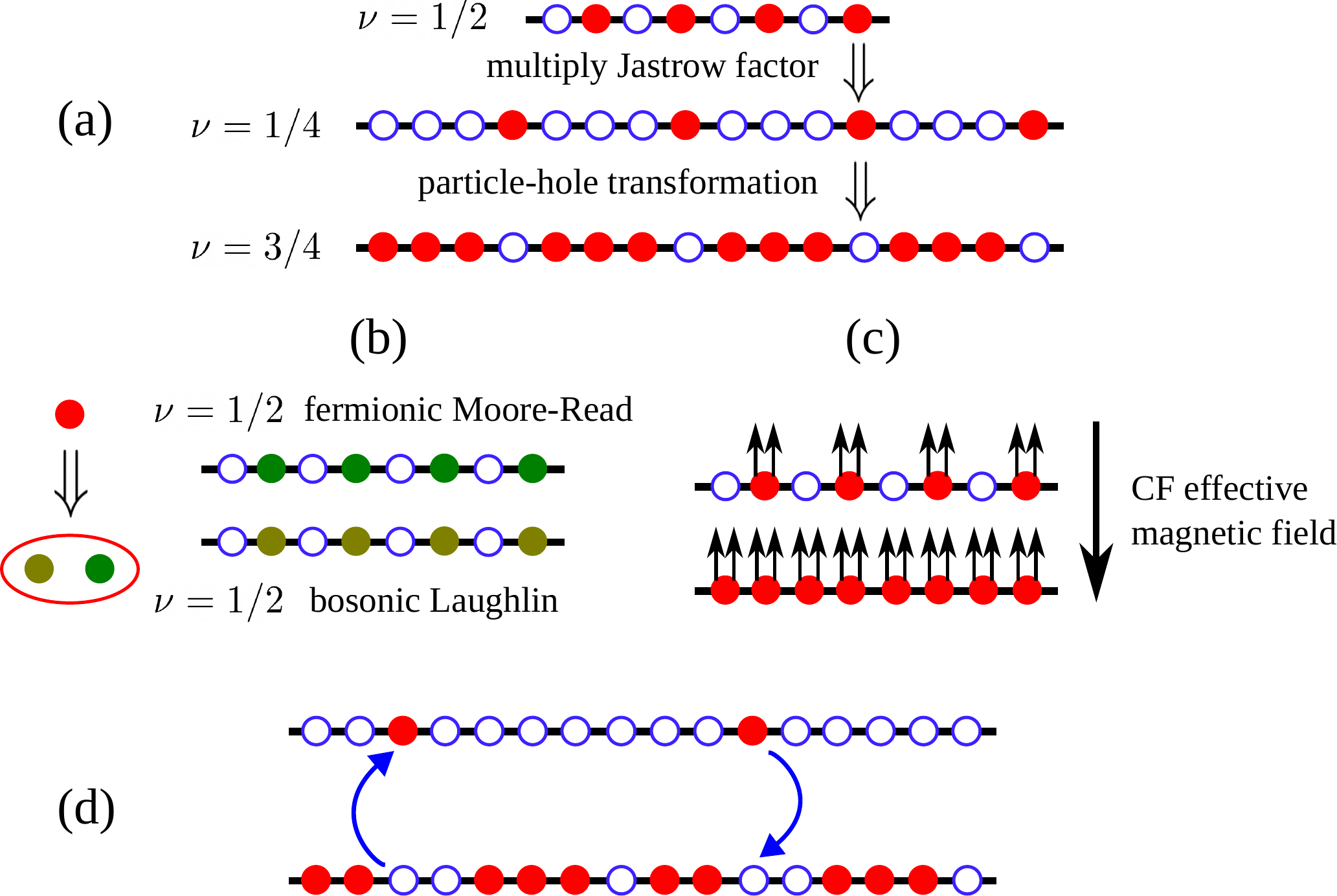}
\caption{(a) Schematics of the $\nu=3/4$ states constructed by the first approach. (b) Parton interpretation of the $\nu=1/4$ states. (c) Schematics of the $\nu=3/4$ states constructed by the second approach. (d) Schematics of the microscopic model used in numerical calculations.}
\label{Figure1}
\end{figure}

The two approaches for constructing non-Abelian FQH states at $\nu=3/4$ are illustrated in Fig.~\ref{Figure1} (a-c). It is helpful to begin with the extensively studied $\nu=1/2$ states. The original Moore-Read Pfaffian wave function written in the disk geometry is
\begin{eqnarray}
{\rm Pf} \left( \frac{1}{z_{j}-z_{k}} \right) \prod_{j<k} (z_{j}-z_{k})^{2}.
\end{eqnarray}
For an isolated LL with two-body interaction, there is an exact particle-hole symmetry that relates filling factors $\nu$ and $1-\nu$. The exact solution of a many-body system at $\nu=1/2$ should be transformed to itself under this symmetry, but the Pfaffian wave function does not satisfy this condition. This fact motivated the construction of the anti-Pfaffian~\cite{Levin2007,LeeSS2007} and the particle-hole symmetric Pfaffian (PHS-Pfaffian) state~\cite{SonDT2015,Zucker2016}. Some wave functions have been proposed for the PH-Pfaffian state~\cite{YangJ2017,Mishmash2018,Rezayi2021} but the anti-Pfaffian state has no simple real space representation. It is possible to construct a parton wave function that has the same universal properties as the anti-Pfaffian state~\cite{Balram2018a}. For our purpose, it is convenient to denote all of them collectively as $\Psi^{\rm MR}_{1/2}$ whenever there is no need to specify the one under investigation. If they are multiplied by another Jastrow factor, we would obtain $\nu=1/4$ states
\begin{eqnarray}
\Psi^{\rm I}_{1/4} = \Psi^{\rm MR}_{1/2} \prod_{j<k} (z_{j}-z_{k})^{2}.
\label{OneQuarterState}
\end{eqnarray}
This wave function can also be interpreted in the parton framework that would be useful in later discussions~\cite{Jain1989b}. In this approach, the same wave function Eq.~\eqref{OneQuarterState} is analyzed but not from the perspective of flux attachment. Instead, one electron is decomposed to two partons, with one being fermionic and the other bosonic. Additionally, emergent gauge fields are introduced to glue one fermionic parton and one bosonic parton to reproduce an electron. The fermionic partons form $\nu=1/2$ Moore-Read type states and the bosonic partons form $\nu=1/2$ Laughlin state~\cite{Laughlin1983}. Its low-energy properties are determined by the gauge fields because the partons are gapped. For each $\nu=1/4$ state, there is a counterpart at $\nu=3/4$ obtained through particle-hole conjugation 
\begin{eqnarray}
\Psi^{\rm I}_{3/4} = \mathcal{PH} \Psi^{\rm I}_{1/4}.
\label{ThreeQuarterState1}
\end{eqnarray}
If the system respects particle-hole symmetry, the states at $\nu=1/4$ and $3/4$ should appear together. Indeed, evidence for a $\nu=1/4$ FQH state in GaAs hole system was reported in Ref.~\cite{WangCY2023a}, but it competes with an insulating state. It is tempting to take them as particle-hole conjugates of each other, but we believe that this identification is premature. As we shall demonstrate below in BLG, gapped state only arise at $\nu=3/4$ when LL mixing is strong. In such cases, $\nu$ and $1-\nu$ are not related by exact particle-hole symmetry. Nevertheless, the universal properties of $3/4$ states can still be analyzed in this way. Monte Carlo calculations have been performed to study trial wave functions at $\nu=1/4$ with Landau level mixing~\cite{ZhaoTZ2023,Sharma2024}. This is not feasible for the $3/4$ state because particle-hole conjugation could not be implemented in coordinate space.

We can also construct $\nu=3/4$ states using the CF theory~\cite{Jain1989a}. An electron binds two fluxes to become CF. FQH states of electrons are interpreted as integer quantum Hall (IQH) states of CFs. This scenario does not work for the $\nu=1/2$ Moore-Read type states, but they can be attributed to CF pairing~\cite{Read2000,Scarola2000a}. To obtain a $\nu=3/4$ state, the CFs should have filling factor $\nu^{*}=-3/2$ according to the relation
\begin{eqnarray}
\frac{\nu^{*}}{2\nu^{*}+1} = 3/4.
\end{eqnarray}
The minus sign means that the CFs form Landau-like levels in negative effective magnetic field as shown in Fig.~\ref{Figure1} (b). An IQH state is realized in the fully occupied lowest level whereas Moore-Read type states appear in the half-filled second level. This analysis leads to the wave function
\begin{eqnarray}
\Psi^{\rm II}_{3/4} = \left[ \Psi^{\rm MR}_{1/2} \otimes \Psi^{\rm IQH}_{1} \right]^{*} \prod_{j<k} (z_{j}-z_{k})^{2},
\label{ThreeQuarterState2}
\end{eqnarray}
where complex conjugate appears due reversed effective magnetic field. A similar picture has been proposed to understand the FQH states at $\nu=3/8$ and $3/10$~\cite{Mukherjee2012,Mukherjee2015}. This approach does not rely on particle-hole transformation but requires the presence of residual interaction between the CFs.

\begin{table*}[ht]
\begin{tabular}{c|cccc}
\hline
\hline
state   &  $N_{e}$ & $2Q$ & $c^{-}$ & gravtion energy and chirality \\
        \cline{2-4}          
\hline
$\frac{3}{4}$-Pf      &  $6N-4$   &  $8N-5$  &  $-1/2$ &  one lower positive, one higher positive \\
$\frac{3}{4}$-aPf     &  $6N$     &  $8N-1$  &  $3/2$  &  one lower negative, one higher positive \\
$\frac{3}{4}$-PHS-Pf  &  $6N-2$   &  $8N-3$  &  $1/2$  &  one lower negative, one lower positive, one higher positive \\
\hline
\hline
\end{tabular}
\caption{Information about the three candidates at $\nu=3/4$.}
\label{Table1}
\end{table*}

The final states generated by these two approaches should have a one-to-one correspondence with each other. When Eq.~\eqref{ThreeQuarterState1} and Eq.~\eqref{ThreeQuarterState2} are said to be equivalent, it does not necessarily mean that they have very high overlaps. Instead, the TOs represented by the two wave functions are the same. This equivalence cannot be proved rigorously, but there are good reasons to believe it. In general, we may ask how to understand FQH states at filling factor larger than $1/2$. It is possible to understand them from particle-hole counterparts at $\nu<1/2$. One can also apply the CF theory directly, which leads to negative effective magnetic field for them. If a state has $\nu=p/(2p-1)$ with $p\in\mathbb{N}^{+}$, the CFs have integer effective filling factors. For other cases such as $\nu=4/5$, the CFs have fractional effective filling factors, so residual interactions between them must be incorporated~\cite{Balram2016b}. As long as the state is Abelian, Chern-Simons theory can be constructed in both approaches, which points to the same TO. It is natural to speculate that non-Abelian states at $\nu>1/2$ given by the two approaches are also equivalent. However, non-Abelian topological field theory is generally much more difficult to study, and we cannot reach the same level of rigor as before.

To substantiate the equivalence, we may study how to define Eq.~\eqref{ThreeQuarterState1} and Eq.~\eqref{ThreeQuarterState2} on the sphere~\cite{Haldane1983c}. In this case, a magnetic monople is placed at the center of the sphere to generated a magnetic field. Let us denote the number of electrons as $N_{e}$ and the number of monople fluxes as $2Q$. In finite systems, we generally have $2Q=N_{e}/\nu-S$. The shift $S$ is an important quantity related to spatial curvature~\cite{WenXG1992b}. For the wave functions $\Psi^{\rm I}_{3/4}$ and $\Psi^{\rm II}_{3/4}$, we need to have a more refined description. If Eq.~\eqref{OneQuarterState} with the Pfaffian factor is used to construct $\Psi^{\rm I}_{3/4}$, the ground state has $N_{e}=6N-4$ and $2Q=8N-5$ with $N\in\mathbb{N}^{+}$. These numbers are reproduced in the other candidate $\Psi^{\rm II}_{3/4}$ if the number of CFs in the second effective level is $2N$. For the other two members, similar connections can also be established. Going beyond the ground state, excitations of these states can also be analyzed. If one flux is added, six quasiholes are created, so the fundamental charge is $e/8$. Another support for the equivalence was gleaned from an inspection of the edge physics~\cite{Kumar2024}. We thus name the three states as $\frac{3}{4}$-Pf, $\frac{3}{4}$-aPf, and $\frac{3}{4}$-PHS-Pf if Eq.~\eqref{ThreeQuarterState1} contains the Pfaffian, anti-Pfaffian, and PHS-Pfaffian wave function, respectively. Table~\ref{Table1} summarizes $N_{e}$ and $2Q$ on the sphere, chiral central charge $c^{-}$ of the edge modes, and graviton properties (see below). 

For a specific system that supports half- or quarter-filled FQH states, considerable efforts are required to determine if they are indeed Moore-Read type states and distinguish between the three members. One fruitful probe is the thermal Hall conductance~\cite{Banerjee2018,Dutta2022b,Melcer2024} that directly reveal the chiral central charge of edge states~\cite{Kitaev2006b}. It is also helpful to analyze the daughter states associated with a particular non-Abelian state~\cite{Levin2009,Yutushui2024a,Zhelton2024}. The existence of daughter states provides indirect yet powerful evidence~\cite{Kumar2024}. Here we study the spectral functions of chiral gravitons. This concept was introduced to describe the long wavelength limit of neutral excitations in FQH states that arise from dynamical oscillation of intrinsic metrics~\cite{YangK2016,LiouSF2019,NguyenDX2021a}. These excitations have spin-2 and definite chiralities that could reveal the nature of FQH states. For some two-flux CF states in GaAs, they have been observed in experiments using circularly polarized light~\cite{LiangJH2024}. It was confirmed that gravitons at $\nu<1/2$ ($\nu>1/2$) has negative (positive) chirality. This is a bulk probe very suitable for studying the $\nu=1/2$ Moore-Read type states~\cite{Haldane2021}. The dominant chirality is negative (positive) for the Pfaffian (anti-Pfaffian) state whereas both chiralities should appear in the PHS-Pfaffian state. In the vicinity of $\nu=1/4$, the graviton physics becomes even more intricate: the CF states at $\nu<1/4$ have two graviton modes with negative chirality, the CF states at $\nu>1/4$ have one graviton mode for both positive and negative chiralities, and the $1/4$ CF liquid has one graviton mode with negative chirality~\cite{NguyenDX2021b,NguyenDX2022,Balram2022a,WangYZ2023}. It has been proposed that four-flux CF states can be recast as parton states such that the higher energy mode corresponds to excitations of bosonic partons~\cite{Balram2022a}. These results shall serve as valuable guidance when graviton modes of the $\nu=3/4$ states are analyzed below.

\section{Numerical Results} 

We proceed to numerical calculations about the $3/4$ state in BLG. It is certainly desirable if an accurate model for BLG is employed, but the essential physics can be unveiled using a simplified model that is more amenable to numerical investigations. The two layers in BLG are labeled as $1,2$ and each one hosts two sublattice sites $A$ and $B$. In the tight-binding description of BLG, each lattice site contributes one state so we have a four-dimensional Hamiltonian~\cite{JungJ2014}. Base on the Slonczewski-Weiss-McClure scheme, four hopping parameters $\gamma_{i}$ ($i=0,1,3,4$) are used. An additional parameter $\delta$ is introduced to characterize the onsite potential of $B_{1}$ and $A_{2}$ sites. In the presence of a displacement field, there is a potential difference between the top and bottom layers. The band structure has two valleys $\mathbf{K}_{\pm}$ and the Landau problem can be studied in each valley. 

It is convenient to define the ladder operators $\widehat{\mathsf{a}}$ and $\widehat{\mathsf{a}}^{\dag}$. The number operator $\widehat{\mathsf{a}}^{\dag}\widehat{\mathsf{a}}$ has eigenstates $|\mathsf{S}_{n}\rangle$ that are non-relativistic Landau orbitals. In the $\mathbf{K}_{+}$ valley, the single-particle Hamiltonian is
\begin{eqnarray}
\begin{bmatrix}
\Delta & \frac{3a\gamma_{0}}{\sqrt{2}\ell_{B}} \widehat{\mathsf{a}} & -\frac{3a\gamma_{4}}{\sqrt{2}\ell_{B}} \widehat{\mathsf{a}} & -\frac{3a\gamma_{3}}{\sqrt{2}\ell_{B}} \widehat{\mathsf{a}}^{\dag} \\
\frac{3a\gamma_{0}}{\sqrt{2}\ell_{B}} \widehat{\mathsf{a}}^{\dag} & \Delta+\delta & \gamma_{1} & -\frac{3a\gamma_{4}}{\sqrt{2}\ell_{B}} \widehat{\mathsf{a}} \\
-\frac{3a\gamma_{4}}{\sqrt{2}\ell_{B}} \widehat{\mathsf{a}}^{\dag} & \gamma_{1} & -\Delta+\delta & \frac{3a\gamma_{0}}{\sqrt{2}\ell_{B}} \widehat{\mathsf{a}} \\
-\frac{3a\gamma_{3}}{\sqrt{2}\ell_{B}} \widehat{\mathsf{a}} & -\frac{3a\gamma_{4}}{\sqrt{2}\ell_{B}} \widehat{\mathsf{a}}^{\dag} & \frac{3a\gamma_{0}}{\sqrt{2}\ell_{B}} \widehat{\mathsf{a}}^{\dag} & -\Delta
\end{bmatrix}
\end{eqnarray}
with $a=0.142$ nm. Based on ab initio calculations, it was found that $\gamma_{0}=2.61$, $\gamma_{1}=0.361$, $\gamma_{2}=0.0$, $\gamma_{3}=0.283$, $\gamma_{4}=0.138$, $\gamma_{5}=0.0$, and $\delta=0.015$ (in units of eV)~\cite{JungJ2014}. The single-particle eigenstates can be written as
\begin{eqnarray}
\begin{bmatrix}
f_{00} |\mathsf{S}_{0}\rangle + f_{01} |\mathsf{S}_{1}\rangle + f_{02} |\mathsf{S}_{2}\rangle + \ldots \\
f_{10} |\mathsf{S}_{0}\rangle + f_{11} |\mathsf{S}_{1}\rangle + f_{12} |\mathsf{S}_{2}\rangle + \ldots \\
f_{20} |\mathsf{S}_{0}\rangle + f_{21} |\mathsf{S}_{1}\rangle + f_{22} |\mathsf{S}_{2}\rangle + \ldots \\
f_{30} |\mathsf{S}_{0}\rangle + f_{31} |\mathsf{S}_{1}\rangle + f_{32} |\mathsf{S}_{2}\rangle + \ldots
\end{bmatrix}.
\end{eqnarray}
Since each row is an infinite summation, the coefficients $f_{in}$ can only be computed numerically using proper truncation. We would like to make some further approximations for two reasons. Firstly, the existence and properties of FQH states are strongly affected by the single-particle wave functions. The parameters $\gamma_{i}$ and $\delta$ may not be perfectly accurate for a specific sample. It would be premature to rule out the existence of a FQH state simply because numerical calculations using these parameters fail to produce this state. As the $3/4$ state only appear in a small regime in the experiment~\cite{Kumar2024}, a more reasonable approach is to search for the parameters that can accomplish this. Secondly, this model does not have full rotational symmetry due to the trigonal wrapping term $\gamma_{3}$. While the existence of FQH states does not require rotational symmetry, it is often more convenient to explore the nature of a state when it does posses this symmetry. 

The filling factor range $[-4,4]$ in BLG is spanned by the zeroth and first LLs from each valley and spin. It is reasonable to assume that the $3/4$ state is spin-valley polarized given the experimental results and previous experience. For moderate magnetic and displacement fields, numerical solutions indicate that the dominant coefficient is $f_{30} \sim 1.0$ in the zeroth LL and $f_{31} \sim [0.90,0.95]$ in the first LL. We thus employ the simplified single-particle eigenstates
\begin{eqnarray}
|\Phi_{0}\rangle = \begin{bmatrix}
0 \\
0 \\
0 \\
|0\rangle 
\end{bmatrix}, \;
|\Phi_{1}\rangle = \begin{bmatrix}
0 \\
c_{0a} |0\rangle \\
c_{0b} |0\rangle \\
c_{1} |1\rangle
\end{bmatrix}
\end{eqnarray}
in our calculations. An energy separation $\hbar\Omega$ between the two LLs is introduced. The electrons interact via the screened Coulomb potential in momentum space
\begin{eqnarray}
V_{\rm sc}(\mathbf{q}) = \frac{e^{2}}{4\pi\varepsilon\ell_{B}} \frac{2\pi\ell_{B}}{|\mathbf{q}|} \tanh(|\mathbf{q}|d)
\end{eqnarray}
with $d$ being the screening distance. Exact diagonalization of the many-body system is carried out on rectangular torus whose two sides have lengths $L_{x}$ and $L_{y}$. The number of magnetic fluxes $N_{\phi}$ enclosed by the torus is quantized according to the relation $L_{x}L_{y}=2\pi\ell^{2}_{B}N_{\phi}$ [$\ell_{B}=\sqrt{\hbar/(eB)}$ is the magnetic length]. We assemble these ingredients to obtain the second quantized many-body Hamiltonian
\begin{widetext}
\begin{eqnarray}
H &=& \hbar\Omega \; C^{\dag}_{1,m} C_{1,m} + \frac{1}{2L_{x}L_{y}} \sum_{\{\alpha_{i}=0,1\}} \sum_{\{m_{i}\}} \sum^{\mathbb{Z}}_{q_{1},q_{2}} V_{\rm sc}(\mathbf{q}) \exp \left[ -\frac{1}{2} |\mathbf{q}|^{2} \ell^{2}_{B} - i\frac{2{\pi}q_{1}}{N_{\phi}} \left( m_{1}-m_{4} \right) \right] \; \nonumber \\
&\times& \widetilde{F}_{\alpha_{1}\alpha_{3}}(-q_{1},-q_{2}) \; \widetilde{F}_{\alpha_{2}\alpha_{4}}(q_{1},q_{2}) \; \widetilde{\delta}_{m_{1},m_{3}-q_{2}} \widetilde{\delta}_{m_{2},m_{4}+q_{2}} \; C^{\dag}_{\alpha_{1}m_{1}} C^{\dag}_{\alpha_{2}m_{2}} C_{\alpha_{4}m_{4}} C_{\alpha_{3}m_{3}}.
\end{eqnarray}
\end{widetext}
Here $C^{\dag}_{{\alpha}m}$ ($C_{{\alpha}m}$) is the creation (annihilation) operator with the first index for the two LLs and the second index for different orbitals within each level and $\widetilde{\delta}_{a,b}$ is a generalized Kronecker function that equals 1 if $a$ mod $N_{\phi}=b$ mod $N_{\phi}$. The momentum $\mathbf{q}$ is expressed using two integers as
\begin{eqnarray}
\mathbf{q} = \frac{2\pi}{L_{x}} q_{1} \mathbf{e}_{x} + \frac{2\pi}{L_{y}} q_{2} \mathbf{e}_{y}
\end{eqnarray}
and the interaction form factors are
\begin{eqnarray}
\widetilde{F}_{00}(q_{1},q_{2}) &=& 1 \nonumber \\
\widetilde{F}_{01}(q_{1},q_{2}) &=& \frac{c_{1}}{\sqrt{2}} \left( \frac{2\pi}{L_{y}} q_{2} , \frac{2\pi}{L_{x}} q_{1} \right) \nonumber \\
\widetilde{F}_{10}(q_{1},q_{2}) &=& \frac{c^{*}_{1}}{\sqrt{2}} \left( -\frac{2\pi}{L_{y}} q_{2} , \frac{2\pi}{L_{x}} q_{1} \right) \nonumber \\
\widetilde{F}_{11}(q_{1},q_{2}) &=& |c_{0a}|^{2} + |c_{0b}|^{2} + |c_{1}|^{2} (1-|\mathbf{q}|^{2}/2).
\end{eqnarray}

\begin{figure}[ht]
\includegraphics[width=0.48\textwidth]{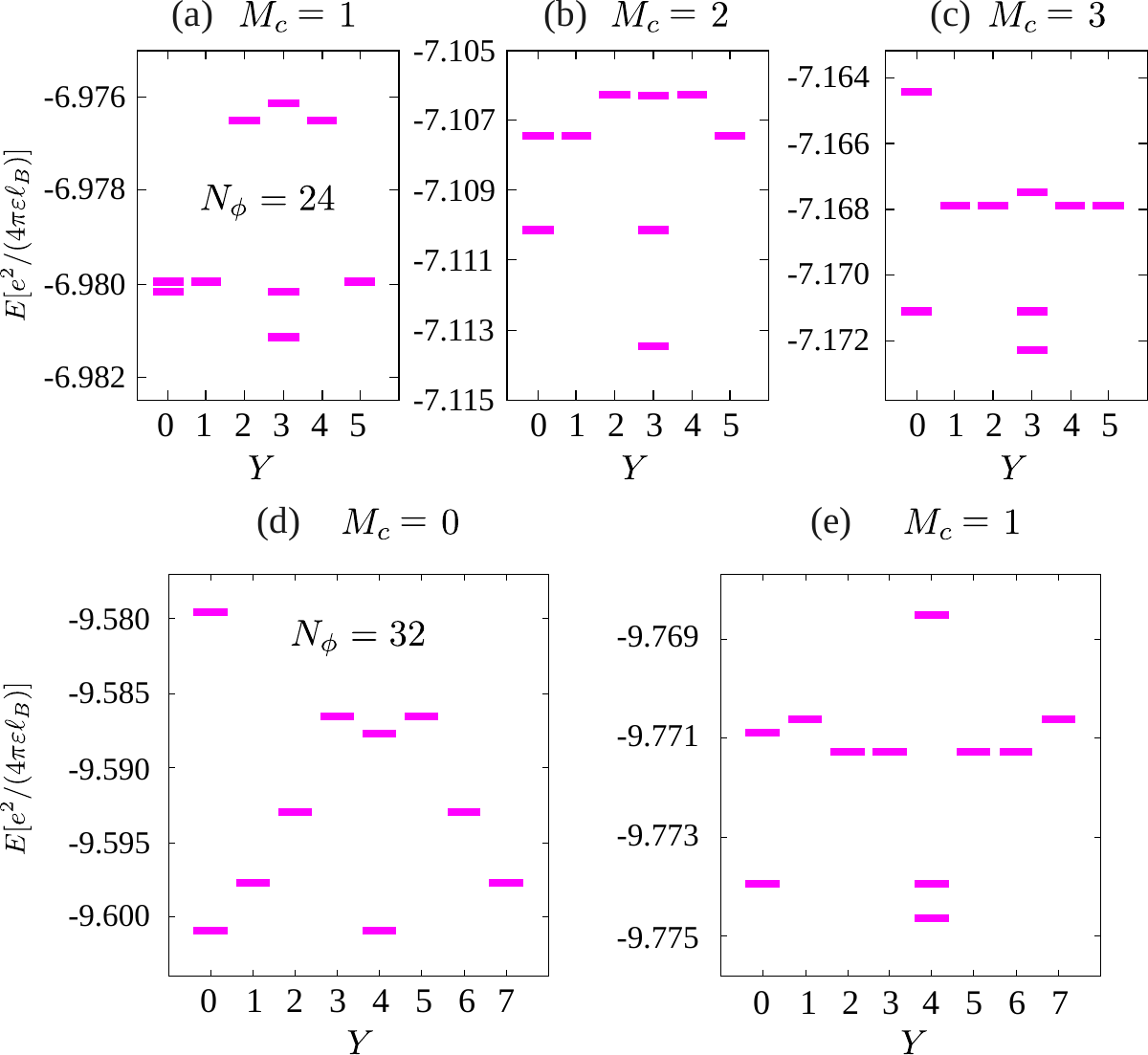}
\caption{(a-c) Energy spectra of the system with $18$ electrons and $N_{\phi}=24$. (d-e) Energy spectra of the system with $24$ electrons and $N_{\phi}=32$. The parameter $M_{c}$ is given in each panel and other parameters are given in the main text.}
\label{Figure2}
\end{figure}

\begin{table}[ht]
\begin{tabular}{c|cccc}
\hline
\hline
system   &  $M_{c}=0$ & $M_{c}=1$ & $M_{c}=2$ & $M_{c}=3$ \\
         \cline{2-4}          
\hline
18 electrons          &  ND     &  4.289  &  1.223  &  0.380  \\
24 electrons          &  ND     &  0.268  &    NA   &    NA   \\
\hline
\hline
\end{tabular}
\caption{The ratios $\epsilon_{\rm sp}/\epsilon_{\rm gap}$ for the energy spectra. Here ND means not defined and NA means not available.}
\label{Table2}
\end{table}

Since the lower level in our model is the same as the non-relativistic lowest LL, no FQH state is expected at $\nu=3/4$ in the absence of LL mixing. It is essential to keep both levels, but then numerical calculations become quite difficult. To this end, an upper bound $M_{c}$ is imposed on the number of electrons that can escape from the lower level to enter the higher one. This strategy has been employed in previous works~\cite{Wojs2006,Rezayi2009}. We choose the parameters to be $\hbar\Omega=0.15 e^{2}/(4\pi\varepsilon\ell_{B})$, $c_{1}=0.95$, and $d=6\ell_{B}$. By summing the momentum $m$ of each particle, we obtain a conserved total momentum $Y\in[0,N_{\phi}-1]$. For a filling factor $\nu=p/q$, the energy spectra in two sectors with $Y$ and $Y+N_{\phi}/q$ are identical~\cite{Haldane1985b}. In other words, there is a trivial $q$ fold degeneracy and it is sufficient to do calculations for $0{\leq}Y<N_{\phi}/q$. The results for $18$ electrons and $N_{\phi}=24$ are displayed in Fig.~\ref{Figure2} (a-c) and those for $24$ electrons and $N_{\phi}=32$ are displayed in Fig.~\ref{Figure2} (d-e). Because the Hilbert space dimension grows very rapidly with $M_{c}$, we can only reach $M_{c}=3$ for 18 electrons and $M_{c}=1$ for 24 electrons. The $3/4$ state is fragile and finite-size effects are quite strong. For $12$ electrons, there is no clear quasi-degeneracy for $M_{c}$ up to $4$. For $18$ electrons, three quasi-degenerate ground states gradually emerge, so the total GSD is $12$. While the desirable degeneracy does not appear at $M_{c}=1$, three states detached from others at $M_{c}=2$, and they get closer when $M_{c}$ increase to $3$. For $24$ electrons, it is quite remarkable that quasi-degeneracy already appear at $M_{c}=1$. To quantify this quasi-degeneracy, Table~\ref{Table2} presents the ratio between two quantities: the energy splitting $\epsilon_{\rm sp}$ within the quasi-degenerate ground states and the energy gap $\epsilon_{\rm gap}$ from the ground states to higher excited states. When the ratio is small, the quasi-degeneracy is good. We note that $\epsilon_{\rm sp}$ cannot be defined properly at $M_{c}=0$ since the second states at $Y=3$ (for $18$ electrons) and $4$ (for $24$ electrons) have very high energies. Based on general bootstrap analysis~\cite{ChengM2025}, this system should support non-Abelian Ising anyons. It is also observed that the $3/4$ state deteriorates as $c_{1}$ decreases in the sense that the GSD is less pronounced. This may be partially responsible for its absence in previous experiments~\cite{Zibrov2017,LiJIA2017a,HuangK2022,HuYW2025} until the most recent one~\cite{Kumar2024}.

\begin{figure}[ht]
\includegraphics[width=0.48\textwidth]{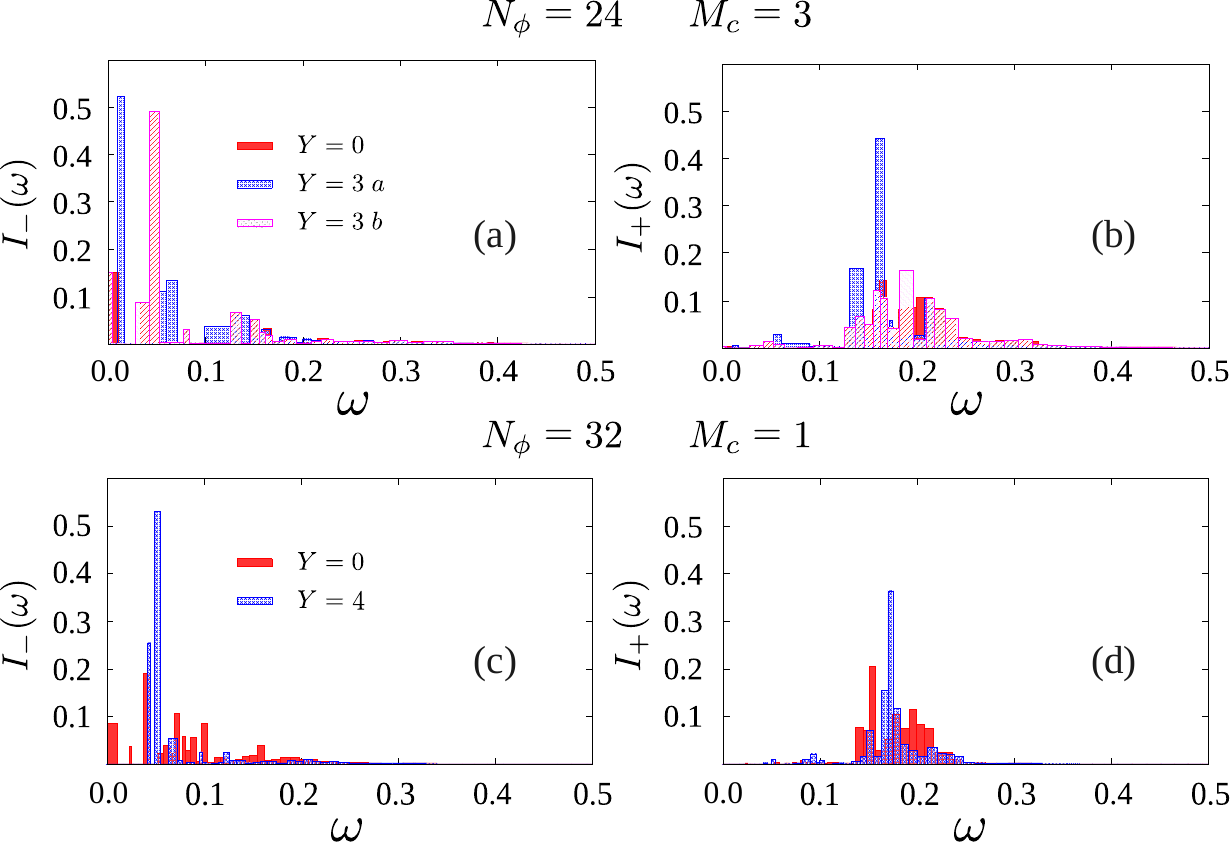}
\caption{(a-b) Chiral graviton spectral functions of the three quasi-degenerate ground states in Fig.~\ref{Figure1} (c). (c-d) Chiral graviton spectral functions of the three quasi-degenerate ground states in Fig.~\ref{Figure1} (e). When there are two states in the same momentum sector, they are distinguished by the label $a$ and $b$. The chirality is negative in panels (a,c) and positive in panels (b,d). }
\label{Figure3}
\end{figure}

The nature of this system can be further probed using the chiral graviton spectral functions. While the $\nu=3/4$ FQH states only emerges in the presence of LL mixing, chiral graviton modes are easier to investigate if the ground states are projected to the lower level. In this case, the chiral graviton operators for the lower level are
\begin{eqnarray}
\mathcal{O}_{\pm} &=& \frac{1}{2L_{x}L_{y}} \sum_{\{m_{i}\}} \sum_{q_{1},q_{2}} \; -\frac{1}{2} \left( \frac{2{\pi}}{L_{x}}q_{1} \mp i \frac{2\pi}{L_{y}} q_{2} \right)^{2} V_{\rm sc}(\mathbf{q}) \nonumber \\
&\times& \exp \left[ - \frac{1}{2} |\mathbf{q}|^{2}\ell^{2}_{B} - i\frac{2{\pi}q_{1}}{N_{\phi}} (m_{1}-m_{4}) \right] \nonumber \\
&\times& \widetilde{\delta}_{m_{1},m_{3}-q_{2}} \widetilde{\delta}_{m_{2},m_{4}+q_{2}} \; C^{\dag}_{0,m_{1}} C^{\dag}_{0,m_{2}} C_{0,m_{4}} C_{0,m_{3}}.
\end{eqnarray}
To derive them, we may perturb the single-particle Hamiltonian using small anisotropic metric~\cite{LiouSF2019} or analyze light-matter coupling using the Luttinger model~\cite{NguyenDX2021a}. In each momentum sector, the many-body eigenstates are denoted as $|\Psi_{n}\rangle$ and the associated eigenstates are $E_{n}$ ($n=0$ is the ground state). Using the Lanczos method of Ref.~\cite{gagliano1987}, we compute the normalized spectral functions
\begin{eqnarray}
I_{\pm}(\omega) = \sum_{n} \frac{\left| \langle \Psi_{n} | \mathcal{O}_{\pm} | \Psi_{0} \rangle \right|^{2}}{W_{\pm}} \delta(\omega - E_{n} + E_{0}).
\end{eqnarray}
with $W_{\pm} = \langle \Psi_{0} | \mathcal{O}^{\dag}_{\pm} \mathcal{O}_{\pm} | \Psi_{0} \rangle$. There is one lower energy mode with negative chirality in Fig.~\ref{Figure2} (d) and one higher energy mode with positive chirality in Fig.~\ref{Figure2} (e). This suggests that the ground states of our model belong to the $\frac{3}{4}$-aPf universality class. As we have mentioned, the $\nu=1/2$ anti-Pfaffian state has one graviton mode with positive chirality. It is expected to be inherited when the $\nu=1/2$ state is converted to $\nu=1/4$. Moreover, a higher energy graviton mode arise from the bosonic partons in Eq.~\eqref{OneQuarterState}, similar to what occurs in the four-flux CF states~\cite{Balram2022a}. When particle-hole conjugation is performed on the $\nu=1/4$ anti-Pfaffian state, the chiralities of both graviton modes are reversed. Based on the same picture, graviton modes in the $\frac{3}{4}$-Pf and $\frac{3}{4}$-PHS-Pf states can also be deduced. The results are summarized in Table~\ref{Table1}. It is also possible to understand these results using the second approach that maps the $3/4$ state to CFs with $\nu^{*}=-3/2$. After the attaching of two fluxes, the CFs in the half-filled second level undergo another flux attachment process to generate $1/2$ Moore-Read type states. This paves the way for two chiral graviton modes.

\section{Discussions} 

In summary, we have studied the possible TOs of $\nu=3/4$ FQH states and revealed that the one observed in BLG is of the $\frac{3}{4}$-aPf type based on numerical results. One may wonder if our results also shed some light on GaAs hole system~\cite{WangCY2022}. It has an intricate LL diagram with many crossings and LL mixing is strong in the vicinity of $\nu=3/4$. The single-particle eigenstates of these LLs are also complicated multi-component vectors. A simple estimate shows that LL mixing is likely to have significant impact and at least two LLs should be included when studying the $3/4$ state. The lower level that is relevant at $3/4$ is also primarily made of the non-relativistic lowest LL orbitals~\cite{MacDonald1989,YangSR1990}. If the higher level can be approximated in the way that we have considered, then the $3/4$ state in GaAs hole system should be essentially the same as the one in BLG. As the stability of the $3/4$ state depends quite sensitively on $c_{1}$, one should perform numerical calculations using microscopic models of GaAs hole system to address this issue.

Besides the $3/4$ state, even-denominator states at $\nu=1/4,3/8,3/10$ have also been found in GaAs hole systems~\cite{WangCY2023a,WangCY2023b}. Since LL mixing plays an essential role at $\nu=3/4$, its existence does not necessarily imply that FQH state also appear at $\nu=1/4$. Indeed, the experiment about BLG did not found a state at $\nu=1/4$~\cite{Kumar2024}, neither do our theoretical model described above. FQH states due to anti-Pfaffian pairing of CFs were predicted at $\nu=3/8,3/10$~\cite{Mukherjee2012,Mukherjee2015}, but they have never been confirmed unambigusouly in GaAs electron systems. It is possible that stronger LL mixing in GaAs hole systems helps to stabilize them~\cite{WangCY2023b}. As for the $\nu=1/4$ state in GaAs hole systems~\cite{WangCY2023a}, more studies are needed to understand the competition between different phases. Further investigations are required to address clarify the nature of these states. One particularly intriguing question is whether different Moore-Read type states could be realized by tuning LL mixing. More generally, many van der Waals materials have LLs that are close in energy. For Bernal stacking trilayer graphene, half-filled FQH states have been observed~\cite{ChenYW2024}, whose properties are determined by details of LL mixing and may evolve with applied external fields. In rhombohedral $N$-layer graphene, there could be $N$ nearly degenerate LLs in certain parameter regimes, which is favorable for realizing some non-Abelian FQH states given by the parton construction~\cite{WuYH2017a,Timmel2023}.

\vspace{1em}

{\bf Note Added} --- Before posting the first version of this manuscript, we became aware of the preprint version of Ref.~\cite{Yutushui2025} that also studies the $3/4$ state. Its method and emphasis are different from ours.

\vspace{1em}

\section*{Acknowledgments} 

We are grateful to Meng Cheng for sharing some results in Ref.~\cite{ChengM2025} before posting it and Hong-Hao Tu for helpful discussions. This work was supported by the NNSF of China under grant No. 12174130.

\bibliography{../../ReferConde}

\end{document}